\documentclass[aps,prb,preprint,onecolumn,groupedaddress,notitlepage,showkeys]{revtex4-1}

  
\usepackage{multirow}
\usepackage{url}
\usepackage{rotating}
\usepackage{graphicx}

\newcommand{\invivo}{\emph{in vivo}}
\newcommand{\invitro}{\emph{in vitro}}
\newcommand{\tcp}{$\mathrm{TCP}_\mathrm{50\%}$}
\newcommand{\ntcp}{$\mathrm{NTCP}_\mathrm{50\%}$}

\newcommand{\ntcpfive}{$\mathrm{NTCP}_\mathrm{5\%}$}
\newcommand{\tcd}{$\mathrm{TCD}_\mathrm{50\%}$}
\newcommand{\ntcd}{$\mathrm{NTCD}_\mathrm{50\%}$}

\newcommand{\gr}{BPNR}

\newcommand{\grfify}{$\mathrm{BPNR}_{50,50}$}
\newcommand{\rgrfify}{$\mathrm{RBPNR}_{50,50}$}

\begin{document} 
\pagestyle{plain}

\def\papertitle{The Biologically Effective Particle Number Ratio (\gr): a new framework to quantify the therapeutic window in SFRT and other modalities}  

\title{\papertitle}
\author{Niels Bassler$^{1,2}$, 
    Giuseppe Schettino$^{3}$, 
    Hugo Palmans$^{3,4}$, 
    Thomas Friedrich$^{5}$, 
    Kelvin Ng Wei Siang$^{6,7}$,
    Emanuele Scifoni$^{8}$, 
    Erik Traneus$^{9}$, 
    Fardous Reaz$^{1,2}$}

\affiliation{$^1$ Department of Clinical Medicine, Aarhus University, Aarhus, Denmark}
\affiliation{$^2$ Danish Centre for Particle Therapy, Aarhus University Hospital, Aarhus, Denmark}
\affiliation{$^3$ National Physical Laboratory, London, UK}
\affiliation{$^4$ MedAustron Ion Therapy Center, Wiener Neustadt, Austria}
\affiliation{$^5$ GSI Helmholtzzentrum für Schwerionenforschung (GSI), Department of Biophysics, Darmstadt, Germany}
\affiliation{$^6$ Erasmus MC Cancer Institute, University Medical Center Rotterdam, Department of Radiotherapy, The Netherlands}
\affiliation{$^7$ Holland Proton Therapy Center, Department of Medical Physics \& Informatics, Delft, The Netherlands}
\affiliation{$^8$ National Institute for Nuclear Physics, Trento Institute for Fundamental Physics and Applications, Trento, Italy}
\affiliation{$^9$ RaySearch Laboratories AB, Stockholm, Sweden}
\date{\today}

\begin{abstract}
\textbf{Background:} Spatially Fractionated Radiation Therapy (SFRT) delivers  highly heterogeneous dose distributions, for which conventional dosimetric metrics (e.g., peak dose, valley dose, average dose, equivalent uniform dose) can give inconsistent or arbitrary estimates of therapeutic window changes, depending on which metric is chosen. 
These quantities are not uniquely linked to biological outcome, making comparisons between modalities problematic.
We introduce the Biologically Effective Particle Number Ratio (BPNR) as a model-independent, outcome-driven framework to quantify therapeutic window changes in SFRT and related modalities.

\textbf{Purpose:} We aim to address these challenges by introducing a robust, model-independent method to quantify changes in the therapeutic window in SFRT. 
The method relies on experimentally accessible quantities and is applicable across treatment modalities.

\textbf{Methods:} BPNR is defined as the ratio of total particle numbers (proportional to monitor units, MU) required to reach specified levels of Tumor Control Probability (TCP) and Normal Tissue Complication Probability (NTCP).
The \emph{relative} BPNR (RBPNR) compares the BPNR values between different treatment modalities. This formulation avoids ambiguities of spatial dose averaging and is grounded directly in measurable biological endpoints.

\textbf{Results:} 
Using preclinical proton minibeam radiotherapy (pMBRT) and conventional proton therapy as test and reference modalities, we derived BPNR values from the TCP and NTCP response curves as a function of MU.
No normal tissue complications were observed in the pMBRT arm at the maximum tested MU, yielding an RBPNR of at least $>$ 1.4 in favor of pMBRT.

\textbf{Conclusions:} The relative BPNR provides a concise, experimentally accessible measure of therapeutic window changes. It is in principle applicable not only to SFRT but also to other emerging modalities such as FLASH, high-LET radiation, and combination therapies. 
This framework complements existing dose-based metrics and enables systematic studies of how spatial, temporal, and biological treatment parameters influence the therapeutic window.
\end{abstract}

\keywords{Spatially Fractionated Radiation Therapy, SFRT, Therapeutic window, BPNR, Radiobiology, Dose heterogeneity, Proton therapy, Dose–response curves}

\maketitle

\clearpage

\section*{Author ORCIDs}
\noindent
Niels Bassler: \url{https://orcid.org/0000-0002-4160-1078}\\
Giuseppe Schettino: \url{https://orcid.org/0000-0003-3284-3953}\\
Hugo Palmans: \url{https://orcid.org/0000-0002-0235-5118}\\
Thomas Friedrich: \url{https://orcid.org/0000-0003-0074-6390}\\
Kelvin Ng Wei Siang: \url{https://orcid.org/0000-0003-4315-4870}\\
Emanuele Scifoni: \url{https://orcid.org/0000-0003-1851-5152}\\
Erik Traneus: \url{https://orcid.org/0000-0002-1850-7382}\\
Fardous Reaz: \url{https://orcid.org/0009-0002-8332-9655}

\section*{Novelty and Significance}
This work:
\begin{itemize}
  \item introduces the Biologically Effective Particle Number Ratio (BPNR), a model-independent, outcome-based metric to quantify therapeutic window changes in SFRT using directly measurable physical quantities;
  \item avoids reliance on peak, valley, or average dose metrics, which can be ambiguous or difficult to interpret in heterogeneous dose distributions;
  \item uses monitor units (MUs)—available with high accuracy and precision—to determine therapeutic window changes for defined biological endpoints (e.g., 50\% TCP and NTCP);
  \item requires no radiobiological modeling, such as LQ-based BED or EUD, making it broadly applicable across modalities and experimental contexts;
  \item enables rigorous and systematic comparisons between modalities, including SFRT, FLASH, high-LET radiation, or combination therapies;
  \item demonstrates feasibility in a preclinical pMBRT setting, supporting translational research and providing a foundation for future large-scale systematic studies.
\end{itemize}

\section{Introduction}
Spatially Fractionated Radiation Therapy (SFRT) is an innovative approach in modern radiation therapy that differs from conventional techniques by delivering radiation in a spatially modulated pattern rather than as a wide uniform field.
This approach produces a characteristic distribution of alternating high-dose ``peaks'' and low-dose ``valleys'' within the irradiated volume, resulting in a heterogeneous dose pattern that has been associated with reduced normal tissue toxicity while maintaining or even improving tumor control compared to conventional treatments~\cite{billena_current_2019, yan_spatially_2020, prezado_divide_2022, prezado_spatially_2024}.
Spatial modulation in SFRT can be achieved using different dose patterns, with spatial features ranging from micrometers in microbeam therapy~\cite{slatkin_microbeam_1992}, to millimeters in minibeam therapy~\cite{dilmanian_interlaced_2006}, and up to centimeter-scale patterns in GRID~\cite{mohiuddin_high-dose_1999} and LATTICE therapy~\cite{wu_modern_2010}.
Each of these techniques produces a non-uniform radiation field with alternating regions of high and low dose, forming the spatially modulated distributions characteristic of SFRT~\cite{prezado_divide_2022}.

Clinical studies of GRID and LATTICE radiotherapy have shown promising results in palliative settings, typically administered as a single fraction with prescribed peak doses in the range of 15 to 20 Gy~\cite{wu_technical_2020, billena_current_2019, yan_spatially_2020}.
Meanwhile, minibeam and microbeam approaches have shown clear potential in preclinical and small-animal studies, with the first clinical case report recently documenting the use of X-ray minibeam therapy in skin cancer patients~\cite{grams_minibeam_2024}.
Although this first-in-human application demonstrates technical feasibility, its clinical advantage over conventional modalities remains to be established: In this case where electron therapy is also an effective treatment option~\cite{ferini_small_2021, yosefof_role_2023}, with feasible and efficient workflows that provide high-quality dose distributions~\cite{canters_clinical_2016, westendorp_automatic_2024}.
A major barrier to a wider implementation is the lack of standardized methodologies to enable systematic investigations into long-term outcomes and optimal treatment parameters.
As a result, there is a need for biologically informed, spatially sensitive frameworks that go beyond simple dose averaging and capture the complexity of heterogeneous SFRT dose distributions.

The biological mechanisms that underlie SFRT’s apparent sparing of normal tissue and preserved tumor control remain poorly understood~\cite{prezado_spatially_2024}. 
Despite encouraging experimental outcomes, there is no consensus on which SFRT parameters most strongly influence biological response~\cite{rivera_conventional_2020, fernandez-palomo_should_2022, chang_journey_2023, prezado_significance_2025}.

To systematically investigate these uncertainties, a robust outcome-driven framework is required to quantify therapeutic window changes in SFRT and related modalities. In this Technical Note we propose such a framework, based on the Biologically Effective Dose Ratio (BEDR).

\subsection{The Therapeutic Window in SFRT}

The \emph{therapeutic window} in radiation therapy refers to the range of exposures within which tumor control is achieved while avoiding unacceptable toxicity in normal tissues. 
Expanding this window is a central goal of technological advances in radiation therapy~\cite{chetty_technology_2015}. 
A related quantity, the \emph{therapeutic index (TI)}, is commonly defined as the tumor response probability at a fixed level of normal tissue toxicity~\cite{benson_therapeutic_2020, joiner_basic_2018}, thereby reflecting the balance between efficacy and side effects. 
An increase in TI corresponds to a \emph{therapeutic gain}, that is, a more favorable ratio of tumor control to normal tissue complications. 

Therapeutic window assessments are often based on comparing the dose required to achieve a given level of tumor control probability (TCP) with that required to reach the same level of normal tissue complication probability (NTCP). 
Although the chosen endpoint is somewhat arbitrary, the 50\% level is commonly used for simplicity and interpretability in preclinical \invivo\ research. 
In clinical contexts, lower NTCP thresholds are typically preferred to ensure acceptable patient risk, while in preclinical studies the availability of full dose–response curves allows greater flexibility in endpoint selection. 

For SFRT, however, this classical dose-based representation becomes problematic. 
Because dose distributions are inherently heterogeneous, the abscissa in such plots (conventionally “uniform dose”) is not uniquely defined. 
Common single-valued metrics such as average, peak, or valley dose conflate fundamentally different spatial patterns and may obscure the biological mechanisms at play. 
Reported average doses also often lack a clear specification of the spatial extent over which they are calculated, whether a single peak-valley pair, a fixed number of modulation periods, or a moving average over a larger volume. 
Because tumor and normal tissue regions are subject to distinct spatial modulations, comparing their respective averages can suggest a therapeutic gain that reflects the averaging method more than a true biological advantage.

Moreover, most SFRT studies report a therapeutic index based on a single prescribed “dose” level, typically derived from such average, peak, or valley dose estimates, rather than from full TCP and NTCP response curves. 
This further limits the interpretability and reproducibility of therapeutic window assessments, since the reported TI depends strongly on the chosen dose metric. 

To address these ambiguities, we introduce the \emph{Biologically Effective Particle Number Ratio (BPNR)}, which quantifies the therapeutic window directly from outcome-based response data, without reliance on spatially averaged dose metrics.
The quantity is based on a very similar concept, which was termed Biologically Effective Dose Ratio (BEDR), which is then generalized to BPNR.

\section{Biologically Effective Dose Ratio}
The \emph{Biologically Effective Dose Ratio} (BEDR) was introduced in the context of antiproton therapy by Holzscheiter et al.~\cite{holzscheiter_biological_2006}. 
It was defined as the ratio of particle fluences required at the beam nozzle to achieve the same biological effect in two different regions of the beam, for example the entrance plateau and the spread-out Bragg peak (SOBP): 

\begin{equation} 
    \label{eq:bedr}
    \mathrm{BEDR} = \left. \frac{\Phi_\mathrm{plateau}}{\Phi_\mathrm{SOBP}} \right|_\mathrm{isoeffect} 
\end{equation}

where $\Phi$ denotes the measured particle fluence at the beam nozzle, and the subscript “isoeffect” indicates that the ratio is evaluated at the same biological endpoint (e.g., 10\% survival \invitro ). 
This definition was motivated by the difficulty of determining absorbed dose reliably in antiproton beams, where dosimetry uncertainties hindered the use of conventional RBE estimates. 
By focusing on fluence ratios under isoeffect conditions, BEDR provided a robust, experimentally accessible way to compare relative biological effectiveness without reliance on uncertain dose measurements. 

A closely related concept was later presented by Grün et al.~\cite{grun_assessment_2015}, who defined the \emph{PER\textsubscript{BIO}} (Particle Effectiveness Ratio) and expressed it in terms of dose and RBE ratios. 
While not identical to BEDR, PER\textsubscript{BIO} highlights the same principle: quantifying biological effectiveness by comparing the input required to achieve matched biological endpoints in different regions or modalities.
PER\textsubscript{BIO} is mathematically similar in form but based on model-predicted RBE-weighted doses at fixed reference positions rather than isoeffect conditions.

\subsection{Generalization to SFRT}

To extend the BEDR concept beyond antiproton therapy and make it applicable to SFRT and other modalities, three adaptations are required. 
First, the spatial reference points are generalized. 
Instead of “plateau” and “SOBP” as used in ion beam studies, we consider biological endpoints in the \emph{PTV} and in one or more \emph{normal tissue assessment regions} outside the PTV. 
These regions may be point-like (e.g. at a certain depth along the beam axis in an \invitro\ setup) or volumetric (e.g. an organ subvolume), and in SFRT their response can vary depending on the local peak–valley contrast along the beam path. 
This generalization avoids modality-specific terminology and ensures that the quantity can be applied wherever tumor control and toxicity endpoints are defined. 

Second, the original BEDR definition required the same biological endpoint in both regions (e.g. 10\% survival \invitro). 
Here we extend this to allow for different but clinically meaningful endpoints, such as TCP in the PTV and \ntcp\ (or \ntcpfive) in normal tissue. 
This extension ensures that the framework is relevant not only for \invitro\ studies but also for preclinical \invivo\ work and, ultimately, clinical translation.

Third, the point quantity fluence is replaced by the total number of particles exiting the treatment nozzle, $N$, a directly measurable and clinically relevant quantity that is proportional to monitor units (MUs) and robustly linked to beam delivery. This yields the \emph{Biologically Effective Particle Number Ratio}, \emph{BPNR}.

With these generalizations, BPNR provides a modality-independent framework for quantifying the therapeutic window on the basis of biological outcomes, rather than spatially averaged dose metrics, independent of the specific model used to fit response curves. 
Here, ‘model-independent’ means independent of mechanistic assumptions about radiobiology. 
BPNR relies only on the endpoints of TCP/NTCP curves, irrespective of whether the curves are interpolated using logistic or probit functions.
What matters is that the chosen fitting approach is transparently reported, so that endpoints are consistently extracted and compared.

\subsection{Operational definition of BPNR}
To implement these adaptations, we define an operational BPNR based on TCP and NTCP response curves:

\begin{equation}
  \label{eq:bedr_operational}
  \mathrm{BPNR}_{50,50} = \frac{N_{\mathrm{NTCP}_{50}}}{N_{\mathrm{TCP}_{50}}} .
\end{equation}

Here, $N_{\mathrm{TCP}_{50}}$ is the total number of particles required to achieve a 50\% probability of tumor control within the PTV, and $N_{\mathrm{NTCP}_{50}}$ is the corresponding number required to reach a 50\% risk of normal tissue complications. 
The 50\% threshold is conventional and convenient for preclinical studies where complete response curves are available, but the definition is flexible: any fixed probability level may be chosen, provided it is applied consistently to both modalities being compared. 
In clinical settings, lower NTCP thresholds (e.g.\ 5–10\%) are typically more relevant, but the same operational framework applies. 

This outcome-based BPNR preserves the iso-effect spirit of the original BEDR definition, complements model-based approaches such as PER\textsubscript{BIO}, and offers a flexible and experimentally robust way to quantify therapeutic differential. 
It can be applied not only to SFRT but also to other biologically modulated techniques such as FLASH, high-LET radiation, or combination therapies.

\subsection{Relative BPNR (RBPNR): comparing modalities}

Once BPNR values are defined for a given treatment plan, they can be used to compare the therapeutic window between modalities.
For this purpose, we introduce the \emph{RBPNR}, defined as the relative change in the operational BPNR values obtained for a test modality and a reference modality: 

\begin{equation}
\label{eq:bedr_ratio}
\mathrm{RBPNR} = \frac{\mathrm{BPNR}_{\mathrm{test}}}{\mathrm{BPNR}_{\mathrm{ref}}}.
\end{equation}

The RBPNR is thus a ratio of ratios and provides a dimensionless measure of how the therapeutic window changes when moving from one treatment technique to another. 
Values $> 1$ indicate a widening of the therapeutic window in the test modality, relative to the reference, while the values $< 1$ indicate a narrowing relative to the reference. 

In practice, BPNR values are extracted from TCP and NTCP response curves obtained by uniformly scaling the monitor units (MUs) of a treatment plan. 
Because MU scales linearly with particle number and, therefore, with dose in all voxels of a plan, the shape of the spatial distribution is preserved, while the biological effect varies systematically. 
This makes MUs a robust and experimentally accessible surrogate for $N$, ensuring that both the TCP and NTCP endpoints are derived under internally consistent conditions. 

\begin{figure}
    \centering
    \includegraphics[width=0.9\linewidth]{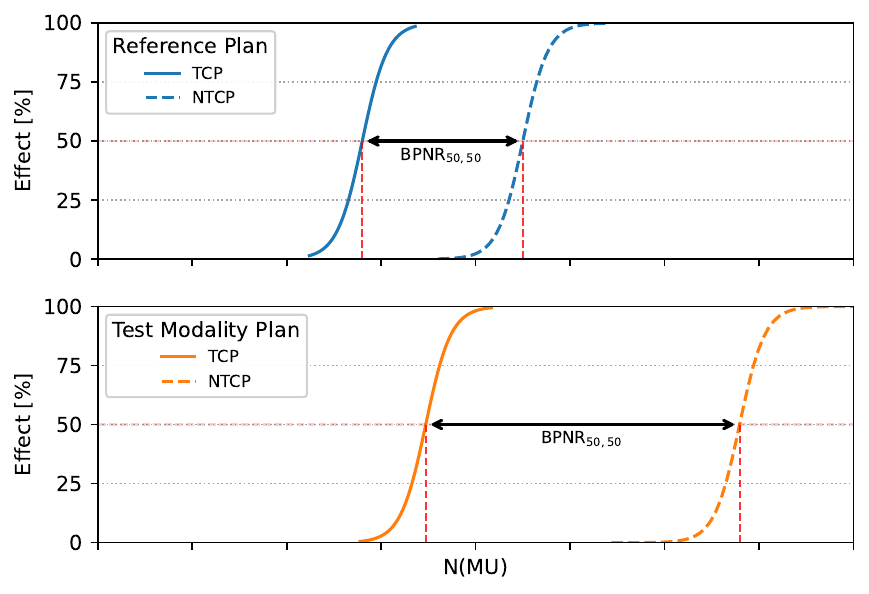}
    \caption{Illustration of the \gr\ concept. 
    Each row represents a treatment modality: a reference plan (top) and a test plan (bottom). 
    For each modality, \ntcp\ and \tcp\ response curves are measured as a function of total particle number $N$ (or equivalently, monitor units, MU). 
    The spatial dose distribution is preserved for each of the two plans, and by scaling the plan MU, full response curves can be established. 
    \grfify is extracted for each plan, and their ratio quantifies the relative change in the therapeutic window between the test and reference modalities. 
    A \rgrfify\ $>$ 1 thus represents an increase in the therapeutic window relative to the reference modality.}
    \label{fig:fig0}
\end{figure}

When comparing two modalities, such as SFRT versus conventional irradiation, each plan is treated independently: MU scaling is applied within each plan, the TCP and NTCP response curves are established, and the corresponding BPNR values are determined. 
The ratio of these values then quantifies the relative therapeutic gain between the modalities. 
An overview of this procedure is illustrated in Fig.~\ref{fig:fig0}, where BPNR values are extracted from response curves for both a reference and a test plan, and their ratio reflects the change in therapeutic window. 

This framework allows systematic evaluation not only of SFRT parameters (e.g.\ beam width, center-to-center spacing, peak–valley contrast), but also of other biologically modulated approaches such as FLASH, high-LET radiation, or combined modality treatments. 
The RBPNR thus provides a unifying outcome-driven metric to assess therapeutic gain across diverse radiotherapy strategies.

\section{Example: application of BPNR to proton minibeam therapy}

To illustrate the application of the BPNR framework, we consider the preclinical \invivo\ experiments with proton minibeam radiotherapy (pMBRT) performed at DCPT~\cite{reaz_probing_2025}. 
In these experiments, tumor bearing and tumor-free legs of mice were irradiated either with a conventional uncollimated proton beam (reference plan) or with a spatially fractionated minibeam configuration (test plan). 
Tumor response (\tcp) and normal tissue toxicity (\ntcp) were recorded as a function of delivered monitor units (MUs), which scale directly with particle number $N$ obtained at the beam nozzle, as long as the treatment plan with the relative spot and layers weights remains unchanged. 
This condition only holds for deriving TCP and NTCP within a given plan, thus, when comparing different modalities (e.g. conventional vs.\ pMBRT), each plan must be evaluated independently to obtain their respective BPNR.

\begin{figure}[htbp]
    \centering
    \includegraphics[width=0.9\linewidth]{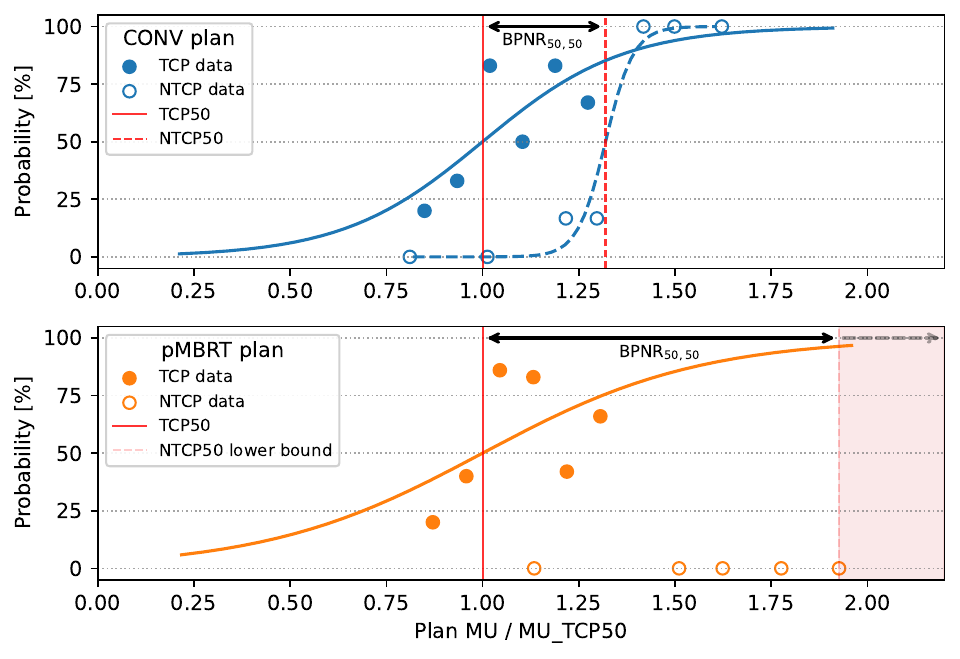}  
    \caption{Illustrative TCP and NTCP response curves for conventional proton therapy (CONV treatment plan, top panel) and proton minibeam radiotherapy (pMBRT treatment plan, bottom panel), shown as a function of delivered monitor units. The abscissa is normalized to the MU needed for achieving ${\mathrm{TCP}_{50}}$ for the respective treatment plan.
    \grfify\ values (black intervals) are derived from the ${\mathrm{TCP}_{50}}$ and ${\mathrm{NTCP}_{50}}$ points for the CONV and the pMBRT plan.
    In the pMBRT case, the absence of normal tissue complications up to the maximum MU provides only a lower bound for $N_{\mathrm{NTCP}_{50}}$ and thus for the \grfify\ (indicated by the grey arrow in the lower panel).}
    \label{fig:example}
\end{figure}

The tissue response is plotted as a function of MU values for both modalities and shown in Fig.~\ref{fig:example}. 
In the conventional arm, both tumor control and normal tissue complications increased with MU, allowing extraction of $N_{\mathrm{TCP}_{50}}$ and $N_{\mathrm{NTCP}_{50}}$ and thereby the corresponding \grfify\ value. 
In the pMBRT arm, no normal tissue complications were observed even at the highest MU level tested.
Even if incomplete, this still yields a lower bound of \grfify\ $> 1.9$ for pMBRT, compared to \grfify\ $\approx 1.3$ for the conventional plan. 
The resulting \rgrfify\ exceeds 1.4, indicating a substantial widening of the therapeutic window under pMBRT delivery. 

\begin{table}[ht]
    \centering
    \caption{Illustrative RBPNR calculation based on the DCPT preclinical dataset~\cite{reaz_probing_2025} (here, using NTCP score 2.5 as described in that reference). 
    Values correspond to the MU levels required to reach 50\% TCP and NTCP in conventional and pMBRT plans, as shown in Fig.~\ref{fig:example}. 
    For pMBRT, no normal tissue complications were observed up to the maximum MU tested, yielding only a lower bound.}
    \label{tab:example}
    \begin{tabular}{l|c|c|c}
        
        \textbf{Plan} & \textbf{non-PTV (NTCP\textsubscript{50})} & \textbf{PTV (TCP\textsubscript{50})} & \textbf{BPNR\textsubscript{50,50}}\\
        \hline
        \hline
        Conventional & 180946 MU & 137136 MU & 1.32 \\
        pMBRT & $>$ 591543 MU & 307068 MU & $>$ 1.93 \\
        \hline
        \multicolumn{4}{r}{\textbf{RBPNR} $>$ 1.46} \\
    \end{tabular}
\end{table}

In Table~\ref{tab:example}, the conventional plan serves as the reference, while the pMBRT plan is evaluated as the test modality. 
\tcd\ and \ntcd\ were extracted from response curves as a function of MUs, for both plans. 
For the pMBRT plan, no normal tissue complications were observed even at the highest MU level tested, so this value is treated as a lower bound on $N_{\mathrm{NTCP}_{50}}$ and consequently for BPNR and the RBPNR. 

In addition to the \grfify\ case illustrated in Table~\ref{tab:example}, the same procedure can be applied to other thresholds. 
For example, using NTCP\(_{5\%}\) and TCP\(_{50\%}\), the resulting RBPNR was $>1.59$ (with a new CONV $\mathrm{BPNR}_{5,50} = 1.21$ while keeping the lower NTCP bound for the pMBRT arm unchanged).

This example demonstrates the practical utility of the RBPNR: it 
quantifies the widening of the therapeutic window directly from experimentally measured outcomes, without reliance on averaged dose metrics or assumptions about peak and valley dose definitions.
This shift from dose-based to delivery-based quantities also points toward broader opportunities for comparing different SFRT configurations, as discussed below.

\section{Discussion}

Previous studies of pMBRT have reported an increased therapeutic index based on a single dose point~\cite{prezado_proton_2018, bertho_first_2021, prezado_spatially_2024}. 
Such definitions are incomplete, as they do not capture the full breadth of the therapeutic window. 
In situations where tumor control and normal-tissue sparing occur in distinctly different dose ranges, a single-point metric cannot adequately represent the true therapeutic advantage.

The RBPNR addresses this limitation by providing an experimentally accessible measure that compares biological effects relative to a reference radiation setup. 
Unlike conventional dose-based descriptors, which are difficult to interpret in the highly heterogeneous fields of SFRT, MU-based RBPNR quantifies therapeutic window changes directly from measured endpoints (e.g.\ TCP and NTCP), without reliance on averaged dose metrics or model-based corrections.

An alternative approach to evaluating SFRT is the Equivalent Uniform Dose (EUD) method~\cite{ahmed_dosimetric_2023}, which condenses heterogeneous dose distributions into a single uniform dose that would produce the same biological response.
It has been found to correlate strongly with tumor response~\cite{rivera_conventional_2020} and has seen widespread application in nonuniform tumor irradiation. 
However, EUD-based methods are based on a priori radiobiological modeling, typically employing the Linear-Quadratic (LQ) model and assumed $\alpha/\beta$ ratios, to translate from \invitro\ cell survival data to \invivo\ therapeutic outcomes.
These assumptions may not always hold, particularly in the high-dose regions characteristic of SFRT.

In contrast to EUD-based methods, which rely on a priori radiobiological assumptions (e.g.\ LQ modeling with fixed $\alpha/\beta$ values), BPNR is derived directly from isoeffect endpoints such as TCP or NTCP. 
These endpoints may be obtained from different sources and fitting procedures, for example from \invitro\ survival curves modeled with LQ or from \invivo\ tumor control and complication data fitted with logistic or probit functions.
The resulting BPNR values will naturally reflect the underlying choice of endpoint and model, and it is therefore essential that the methodology be reported transparently. 
The strength of the framework is that it is agnostic to the specific model form: once the endpoint is defined, the BPNR calculation itself is a straightforward ratio of measured quantities. 
This flexibility allows BPNR to provide relative measures in \invitro\ contexts while also supporting more direct quantification in \invivo\ studies, thereby facilitating systematic comparison across different modalities and experimental conditions.

In addition, the ability to compare BPNR values across different treatment modalities provides a useful tool for therapy optimization, enabling direct experimental evaluation of how SFRT parameters such as peak dose, valley dose, beamlet width, and spatial fractionation patterns influence the therapeutic window.

An ongoing challenge in SFRT research is the pursuit of a single universal dosimetric metric that predicts biological effectiveness.
To date, no such metric has proven universally applicable, as none fully captures the complex biological effects associated with the spatially heterogeneous dose distributions of SFRT.
The peak dose, valley dose, average dose, EUD, and other dosimetric descriptors remain valuable for reporting purposes~\cite{rivera_conventional_2020, fernandez-palomo_should_2022, prezado_significance_2025}.
However, these parameters should ultimately be considered as supplementary metrics rather than primary metrics to assess relative therapeutic effectiveness.
Reducing SFRT’s complex dose distributions to a few scalar metrics risks losing spatial and biological information that could later prove biologically or clinically important.
A meaningful interpretation and reanalysis require, ideally, that the full voxelized dose distribution be preserved and reported.
Only in this high-resolution form does the dose retain the spatial context necessary for robust modeling and biological interpretation.

Unlike peak or valley dose, which are purely physical descriptors and may give divergent impressions of therapeutic gain depending on which quantity is chosen, BPNR is anchored in biological endpoints. 
Once TCP or NTCP thresholds are specified, the BPNR value is uniquely determined from the measured response data. 
This outcome-based definition reduces the arbitrariness inherent to peak/valley metrics and provides a reproducible basis for comparing different modalities.

While the RBPNR provides a concise, outcome-based summary, it is fundamentally a descriptive measure: it quantifies the relative widening of the therapeutic window between a test and a reference configuration, based on observed TCP or NTCP endpoints. 
Its value therefore depends directly on biologically relevant outcome data obtained from experiment or observation. 
On its own, BPNR does not predict the performance of a new treatment plan. 
Predictive use requires an additional modeling step that relates plan-specific characteristics (e.g.\ voxelized 3D dose distributions or derived dosimetric descriptors) to the expected biological endpoints. 
In such a framework, BPNR provides the outcome-driven metric, while the model supplies the predictive mapping—for example, allowing \ntcpfive\ to be estimated once \tcp\ is 
known for a given plan. 

The strength of the BPNR framework lies in its ability to enable robust, direct comparisons of different SFRT parameters, such as beamlet width, peak–valley spacing, or spatial configuration, by quantifying their relative impact on the therapeutic window independently of specific radiobiological assumptions.
Furthermore, the same methodology can be extended to classical radiotherapy concepts such as the volume effect and temporal fractionation, as well as to emerging approaches including FLASH and multimodal combination therapies~\cite{schneider_combining_2022}.

\section{Conclusion}
Quantifying the therapeutic window in SFRT remains a significant challenge due to the spatial complexity of dose distributions and the limitations of conventional metrics such as average dose or EUD. 
In this work, we extend the BPNR concept to define a model-independent, outcome-driven measure of therapeutic gain that is compatible with heterogeneous fields and clinically relevant endpoints such as \tcp\ and \ntcpfive.
By relying on the ratio of total particle numbers, proportional to monitor units, required to reach defined biological outcomes, the BPNR and its derivative RBPNR offer a practical and experimentally grounded framework for comparing different treatment modalities. 
Most importantly, this approach enables systematic investigations of how parameters such as spatial fractionation, peak–valley contrast, and beamlet geometry influence the therapeutic window. 
Moreover, the framework naturally enables comparisons across both conventional and biologically modulated therapies, including FLASH, high-LET, and hyperthermia-enhanced radiation.

\section*{Acknowledgements}
We acknowledge the support of the NovoNordisk Foundation (grant number NNF195A0059372), DCCC Radiotherapy - The Danish National Research Center for Radiotherapy WP10, and the Danish Cancer Society (grant number R191-A11526 and R374-A22657).

\section*{Competing interests}
The authors declare that they have no competing interests.

\bibliographystyle{unsrt}
\bibliography{references}

\end{document}